\documentclass[aps,jcp,twocolumn,groupedaddress]{revtex4-1}

\usepackage{graphicx}

\begin{document}

%\preprint{}

\title{Sampling Rare Events in Non-Equilibrium and Non-Stationary Systems}


\author{Joshua T. Berryman}
\email[]{josh.berryman@uni.lu}
\author{Tanja Schilling}
\affiliation{Johannes Gutenberg Universit\"at, Mainz, Germany}
\affiliation{Universit\'e du Luxembourg, Luxembourg}
%\thanks{}

%\date{\today}

\begin{abstract}
Although many computational methods for rare event sampling exist, this type of calculation is not usually practical for general nonequilibrium conditions, with macroscopically irreversible dynamics and away from both stationary and metastable states. A novel method for calculating the time-series of the probability of a rare event is presented which is designed for these conditions.  The method is validated for the cases of the Glauber-Ising model under time-varying shear flow, the Kawasaki-Ising model after a quench into the region between nucleation dominated and spinodal decomposition dominated phase change dynamics, and the parallel open asymmetric exclusion process (p-o ASEP).  The method requires a subdivision of the phase space of the system: it is benchmarked and found to scale well for increasingly fine subdivisions, meaning that it can be applied without detailed foreknowledge of the physically important reaction pathways.
\end{abstract}



\maketitle

\section{Introduction}

Events which are highly improbable often have great
importance to the behaviour of a system. The classic example of this is
nucleation of a raindrop from supersaturated water vapour. Because droplets
smaller than a critical radius are energetically unfavourable, the formation
of a super-critical droplet is unlikely to occur on the timescale of 
thermal motion of water molecules. Hence a straightforward computer 
simulation would waste much CPU time on unimportant fluctuations before producing 
an event of interest.

\subsection{The Rare Event Literature}

A large number of approaches have been developed to solve so-called ``rare event'' problems.  Many of these approaches are based on Transition State Theory \cite{Eyring1935, Wigner1938}, i.e.~on the concept of a quasi-equilibrium free energy landscape and of a particular ``slow'' motion of the system within this landscape.  The landscape is imagined as consisting of basins linked together by `Transition Paths' passing over saddle points.  This landscape can be mapped using equilbrium methods including e.g.~umbrella sampling \cite{Torrie1977}, multi-canonical sampling \cite{Berg1992} and Wang-Landau sampling \cite{Wang2001}; and the motion across saddle points, once they are identified, can then be simulated directly by initialising molecular dynamics simulations near to the saddle or even just reconstructed from the potential of mean force \cite{Strodel2007}.  This approach of proceeding from a free energy surface  to an understanding of the kinetics is principled and highly attractive but is limited to systems for which such a surface can be meaningfully defined and practically computed.  

For systems away from equilibrium the concept of free energy becomes problematic, although theories which use analogues or extensions of this idea are rapidly being developed; for example by considering the transition probabilities between states \cite{Evans2010} rather than directly assigning a free energy and the associated Boltzmann probability distribution directly to states themselves.

In order to numerically study rare events under conditions where the concept of a free energy landscape is problematic, Transition Path Sampling (TPS) \cite{Dellago1998}, Forward Flux Sampling (FFS) \cite{Allen2006, Allen2006a}, Weighted Ensemble (WE) \cite{Huber1996} and a suite of related methods have been developed.  The basic idea of these methods is to selectively sample from the set of pathways which the system can take, by increasing the number of pathways in the important regions of the state space of the system, but compensating by attaching a variable statistical weight to each path. This group of methods is aimed at steady state non-equilibrium systems, and except for one very recent paper on WE \cite{Zhang2010} (published after this work was substantially complete) the potential for adapting or reformulating them to give a time-dependent description of non-stationary dynamics has not yet been explored. Many processes (such as quenching, aging, ignition and impact) are naturally framed in a strictly non-steady or time-dependent way: beyond the equilibrium/quasi-equilibrium and also the stationary nonequilibrium treatments.  The development of non-stationary rare event methods is therefore of potentially great importance.

\subsection{Phase-Space Binning and Reweighting}

The starting point for this work is a phase space binning according to some macroscopic coordinate $\lambda$ (which is often called the ``reaction coordinate'' although it is not usually the true reaction coordinate of the process). Biased sampling is then performed so as to generate paths which move through specific bins on $\lambda$.  This strategy of projecting the phase space of the system onto some subspace of one or more dimensions; dividing the subspace with a set of partitions and then running short trajectory paths in or between compartments is common to FFS \cite{Allen2006}, WE \cite{Huber1996}, Milestoning \cite{Faradjian2004} and Boxed Molecular Dynamics \cite{Glowacki2009} and has been very successful.  Although these algorithms do not all require detailed balance and are successful for treating non-equilibrium steady states, application to general non-stationary dynamics is still exploratory, and is so far only shown for WE \cite{Zhang2010} (although the authors step back from actually claiming this, preferring to state that the method covers a `broad class of stochastic processes' rather than the full range of stochastic non-stationary dynamics). The requirement for stationarity in all existing methods apart from WE arises from the assumption that microstates from a given compartment can be treated interchangably at the compartment boundaries - regardless of, often importantly, the duration of the path which has led to a given state.  

We generalise this strategy of compartmentation to non-steady-state systems; in essence only by fixing the duration of each trajectory fragment (here called a `shot'), so that the time evolution can easily be tracked. The resulting method, once sampling and reweighting schemes have been developed around this central premise, is termed Stochastic Process Rare Event Sampling (S-PRES).  The most important design choice is the procedure used to ensure dynamically adaptive sampling rates for the different bins.  This is achieved here using a variant of Rosenbluth sampling, as is sometimes used in FFS; rather than by moving the bins as has been investigated for WE.  The choice to keep the bin positions fixed has the benefit of allowing a high-level and mathematically friendly description of the dynamics to be developed online in the form of a time-dependent matrix of transition frequencies between the bins.

\section{S-PRES: Algorithm Description}

\subsection{Overview}
\label{sec:overview}
We define a scalar-valued coordinate $\lambda$ as a function over the state space of our system. This coordinate  is discretized into bins labelled by an index $i$. (Note that the phase space of the system need neither be discretized nor finite. These conditions only need to hold for the coordinate binning.) As an example choice for $\lambda$, one might use the number of particles in a liquid droplet forming in supersaturated vapour. 

The main goal of S-PRES is to observe a roughly constant number of forward transitions from each bin $i$ on each interval $[t, t+\tau]$; where a forward transition from $i$ at $t$ is defined as any shot where the microstate at time $t+\tau$ falls within a bin $j>i$. In this way unlikely transitions can be explored and sampled with high statistical accuracy.

In principle, $\lambda$ could be a vector instead of a scalar.  The restriction to a scalar is used here to simplify the discussion.  If using a vector-valued $\vec{\lambda}$, the concept of `forward' becomes non-obvious.  An example approach in two or more dimensions is to define a Hamming distance as the number of bin boundaries which remain to be be crossed in order to reach some target bin: `forward' then describes any shot which decreases this Hamming distance.

\subsection{Importance Sampling for Adaptivity}

We carry out a variable number ${n}^t_{i}$ of `shots' (short dynamics runs) of a fixed duration $\tau$ from the configurations in each bin $i$ at each time $t$.
(Here and in the following we use the term ``configuration'' for a microstate of the system at a given time on a given path. Two paths can, in principle, 
reach identical microstates at the same time. In this case the algorithm would still hold two configurations.)

 As we are considering stochastic processes, shots from the same starting configuration can be made to diverge by varying the random number seed used to generate the dynamics. ${n}^t_{i}$ is adapted during the simulation to improve statistics. During the course of the simulation, the number of bins which are populated by configurations gradually increases (indicated by the triangles and squares in fig.~\ref{fig:reweighting}) until transitions are sampled at each timestep from all bins which have a non-zero occupation probability. In order to achieve a roughly constant number of forward transitions from each bin $i$, we  select $n^t_{i}$ based on the estimated transition probabilities at the previous timestep.  The number of shots from bin $i$ at time $t+\tau$ is defined as:

\begin{equation}
\label{eqn:sampfreq}
n^{t+\tau}_i = \lceil n^t_{i} + \gamma \left( \frac{N}{R} -1 \right) n^t_{i} \rceil 
\end{equation}

where $N$ is the target number of forward transitions, $\gamma$ is a damping factor, $R$ is the number of shots which moved forward from bin $i$ to bins over higher ranges of $\lambda$ on the step $t-\tau$ to $t$, (or $R = 1$ if this number is zero). The brackets $\lceil \mathrm{~} \rceil$ indicate the ceiling function. This adaptive sampling method makes S-PRES akin of the class of variational approaches to steady-state importance sampling (IS) described elsewhere \cite{Cai2002}.  Selection of values for the parameters $N$ and $\gamma$ is discussed in sec.~\ref{sec:params}.

\subsection{Explanation of Sampling for Path Generation}

In order to enhance the exploration of rare states we apply a version 
of the pruned-enriched Rosenbluth method (PERM) \cite{Grassberger1997}, 
i.e.~when picking configurations from a given bin as starting points for new 
shots, we do not select them with equal probabilities, but with a variable statistical weight which depends on the sampling history.  In conventional PERM paths at a given timepoint are either discarded, or selected for exactly one or two copies to extend to the next timepoint, based on lower and upper thresholds in their weights.  The implementation presented here avoids having to set these thresholds. The number of branches $n^{t}_i$ is shared out between paths in each bin $i$ randomly in proportion to the path weights.  This has the same effect on average of discarding the relatively unlikely paths and sharing out the weights of the relatively more likely paths; such that statistics over a given bin are not dominated by a few highly weighted paths and also such that computational effort is not wasted on highly unlikely paths which contribute almost nothing to the statistics.

\begin{figure}
\includegraphics[width=0.5\textwidth]{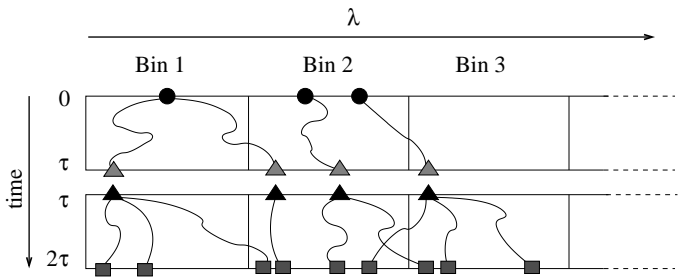}
\caption{\label{fig:reweighting} {\bf Schematic representation of path 
generation}.  Symbols indicate configurations, lines represent path 
segments of duration $\tau$.}
\end{figure}

Here we give a non-mathematical introduction to the procedure (see fig.~\ref{fig:reweighting}, where 
symbols stand for configurations and lines indicate path segments) and then 
we motivate the method further in \ref{sub:Mot}.   

Initially, 
configurations in each bin $i$ with non-zero number of occupying configurations are selected with 
equal probability as starting points for the $n^0_i$ paths from that bin 
(in the example of fig.~\ref{fig:reweighting} with probability 1 for the single circle in bin 1 at time=0; and probability 1/2 for each of the two circles in bin 2 at time=0). Subsequently, we take into 
account from which bin $i$ a configuration in $j$ originates (the number 
of branches extending from the configurations in $i$ determines the 
weight of their endpoints in $j$.) 
In fig.~\ref{fig:reweighting}, the 
left triangle in bin 2 stems from a path with weight 1/2 (1 state divided by 
2 branches), while the right triangle stems from a path with weight 1 
(2 states by 2 branches). Hence, when selecting starting configurations for new shots, the left triangle in bin 2 is chosen with half the probability of the right triangle.

\subsection{Motivation of Sampling Strategy}
\label{sub:Mot}

We bias the sampling of trajectories towards rare events by adapting the number of shots  $n^t_i$ from a bin $i$ at a time $t$ such that sufficient statistics are produced for rare transitions (eqn.~\ref{eqn:sampfreq}). However, we do not bias {\it within} a bin: when we select a sample of $n^t_i$ configurations from the path endpoints in a given bin $i$ as starting points for new shots, we do not apply any bias.

Selecting configurations without bias does not imply that they are drawn with equal probabilities. On the contrary, as the total number of shots $n^t_i$ varies betwen bins, pathways that arrive in a bin $j$ from different bins $i, j, k \ldots$  have different statistical weights (according to the respective values of $n^{t-\tau}_i, n^{t-\tau}_j, n^{t-\tau}_k \ldots$ ). These weights need to be taken into account when selecting starting configurations for new shots. We now provide a detailed explanation of the procedure to do this.

We define $P^t(i)$ as the proportion of configurations in bin $i$ at time $t$ assuming infinitely many configurations. We call the estimate of $P^t(i)$ from a finite number of configurations $d_i^t$. Similarly, 
we define $P^t(j|i)$ as the proportion of pathways from bin $i$ which end in $j$ in the 
case of infinitely many trajectories, and its estimate as $M_{i,j}^t$. 

As in the example of fig.~\ref{fig:reweighting}, we begin at time $t=0$ by picking 
configurations in bin $i$ with equal probability as starting 
configurations for shots. At time $t=\tau$ we count $N^\tau_{i,j}$ pathways that went from $i$ to $j$. Each of these pathways has an equal weight $P^\tau(i,j)/N^\tau_{i,j}$, because they each had the same chance to be selected for shots from bin $i$. For the next step, we would like to pick configurations from bin $j$ such that the probability of a configuration $s_j^\tau$ being picked is proportional to the weight of its path $P(\mathrm{select}: s_j^\tau) \propto P^\tau(i,j)/N^\tau_{i,j}$. To conveniently compute this we normalize by $P(j)$ and write:

\begin{eqnarray}
P(\mathrm{select}: s_j^\tau) & = & (\frac{1}{N^\tau_{i,j}})P^\tau(i,j)/P^\tau(j)\nonumber\\
P(\mathrm{select}: s_j^\tau) & = & (\frac{1}{N^\tau_{i,j}})P^\tau(j|i)P^\tau(i)/\sum_{i'}P^\tau(j|i')P^\tau(i')\nonumber
\end{eqnarray}

During the course of the simulation we do not know the values 
of $P^\tau(j|i)$ and $P^\tau(i)$. However, we do have the estimates
$M_{i,j}^\tau$ and $d_i^\tau$. As we sample within a bin without bias, 
the errors in $d_i^\tau$ (and $M_{i,j}^\tau$) relative to $P^\tau(i)$ 
(and $P^\tau(j|i)$) are zero-mean. Therefore all products and ratios of different errors 
are also zero-mean. And hence the laws of conditional probability 
can be applied to the estimated probabilities without introducing any bias. Therefore an estimate of the optimal $P(\mathrm{select}: s_j^\tau)$ can be defined as a function of $M_{i,j}^\tau$ and $d_i^\tau$, without introducing any bias:

\begin{eqnarray}
\label{eqn:psel}
\overline{P}(\mathrm{select}: s_j^\tau) & = &  \frac{M^\tau_{i,j}d^\tau_i} {N_{i,j}\sum_{i'}M^\tau_{i',j}d^\tau_{i'}}
\end{eqnarray}

Selecting configurations for shots using $\overline{P}(\mathrm{select})$ instead of $P(\mathrm{select})$ is perfectly acceptable: in a large number of repeated experiments each configuration will be selected a number of times proportional to its true $P(\mathrm{select})$, and correct average properties will be observed.

The central trick of the algorithm is that although the trajectories which 
enter a given bin do not have an equal statistical weight; those which leave 
a given bin {\it do} have an equal statistical weight, because their selection for shots is determined by an unbiased estimate of $P(\mathrm{select})$.  For this reason, the selection formula (\ref{eqn:psel}) can be applied at every timestep without explicitly considering the histories of trajectories more than one timestep into the past.

\subsection{Algorithm Parameters}
\label{sec:params}

The duration of path segments $\tau$, the `damping factor' $\gamma$ (from eqn.~\ref{eqn:sampfreq}) and the target number of forward transitions per bin $N$ must be set by the practitioner.  For the method to be efficient, $\tau$ should be shorter than the shortest passage time of the rare event and longer than the timescale necessary to move from one $\lambda$-bin to another. If the dynamics can vary such that the time required to transit between $\lambda$-bins is often longer than $\tau$, then a small value of $\gamma$ (say between $0.01$ and $0.5$) can improve stability by stopping the adaptive sampling from attempting too many shots before it is possible that even one of them moves forward. On the other hand, in the case that $\tau$ should come near to the timescale over which the dynamics can evolve globally with respect to time then $\gamma$ should be set close to $1$, allowing the sampling rate to adjust rapidly.

 It is possible to dynamically vary both $\gamma$ and $\tau$ if needed; for instance in a glassy material $\tau$ can be increased as the system arrests.  If the system has a complex configurational space, with multiple substantially different reaction pathways all projecting onto the same values of $\lambda$, then `dead-end' states may sometimes be reached, where $n_i$ grows out of control even for small $\gamma$ and large $\tau$.  In this case, if a better $\lambda$ is not available, a threshold should be defined for the largest acceptable $n_i$ given practical computational constraints.  $N$ should be set large enough such that it is larger than the typical fluctuations in $n_i$.

\subsection{Extraction of Observables and Statistics}
\label{sec:observables}
In order to compute expectation values of observables, we consider time slices throught the set of pathways. A configuration $s$ from the set of configurations at time $t$ is associated with a microscopic weight $w^t_s$. All configurations at the same timepoint in an S-PRES run have had the same history of external control parameters. The microscopic weight $w^t_s$ of a configuration $s$ is given by:

\begin{eqnarray}
\label{eqn:microweight}
 w^t_s & = &  \overline{P}(\mathrm{select}: s_j^t)d^t_j\nonumber\\
 w^t_s & = &  \frac{M^t_{i,j}d^t_i} {N_{i,j}}
\end{eqnarray}

 using the same reasoning as eqn.~\ref{eqn:psel}. $w^t_s$ is proportional to the estimated probability of occurrence of the configuration at the same timepoint in a repeated experiment with the same control parameter history and distribution of initial configurations (but, obviously, with different individual trajectories due to the stochastic nature of the dynamics).

In order to extract the expected value of some observable $x^t$ it is required to take an average over all configurations $s$ at time $t$.  The sum of the weights over all configurations is $1$. Then the time slice average is given by:

\begin{eqnarray}
\label{eqn:obsweights}
\overline{x}^t             & = & \sum_{\{s^t\}} w_s^t x(s^t)
\end{eqnarray}

This sample average will converge with a sufficiently large number of configurations.  However, estimates of fluctuations require some care because the different configurations held at a given time are likely to have some degree of mutual information, having previously branched from a single parent or grandparent configuration, leading to potentially dramatic underestimates of the variance.  The simplest solution to this problem is to run two completely independent calculations from different sets of starting configurations, calculating the variance at $t$ in the second run based on the estimated mean at $t$ from the first run (and vice-versa).  Estimates of the error (as distinct from fluctuation) either in direct observables or in variances can then be achieved by making further independent runs.

\subsection{Macroscopic Description of Time-Evolution}
\label{sec:macro}
In order to estimate the progress of the algorithm online, we consider the 
system dynamics in terms of the macroscopic coordinate $\lambda$. These 
dynamics will in general not be Markovian. For the following discussion, 
we borrow (with apologies) some mathematical notation from the language 
of Markov processes, but we do not imply that the dynamics in terms of 
$\lambda$ are Markovian. 

We consider a time series of `macrostate vectors' 
${\vec d}^t$ of the estimated occupation probabilities of the bins, beginning 
from the initial distribution which has been chosen and progressing such 
that: ${\vec d}^{(t+\tau)} = ({\vec d}^t)^{\mathrm T}{\mathbf M}^t$.  

We can define the entries of ${\vec d}^t$ and ${\mathbf M}^t$ just like any other observables, as the sum of the associated weights (as from eqn.~\ref{eqn:microweight}):

\begin{eqnarray}
\label{eqn:markov}
{d^t_j}    & = & \sum_{\{s^t\}} w_s^t \delta^t_j  \\
{M^t_{i,j}} & = & \sum_{\{s^t\}} w_s^t \delta^{t-\tau}_i \delta^t_j   / d^{t-\tau}_i
\end{eqnarray}

where $\delta^t_i$ is $1$ if the path occupies bin $i$ at time $t$, otherwise $0$.

 In the case of a Markov process, a marginalisation is carried out over `parent' bins $i$ such that ${d^t_j} = \sum_i d^{t-\tau}_iM^{t-\tau}_{i,j}$.  Because we have time-varying control parameters, and because we do not make the assumption of memorylessness, this marginalisation acquires caveats. Obviously, we must bear in mind that the estimated occupation probabilities depend on the history of control parameters; and that they are by definition time-dependent.  Less obviously, it does not hold that the marginalisation over the `parent' bins $i$ can be carried out without loss of information as in the case of Markov dynamics, meaning that our $M^{t-\tau}_{i,j}$ represents an average over the configurations in $i$, rather than having the same value for each configuration in $i$. In particular, the practitioner should be aware that non-trivial correlations of the process on a time-scale longer than $\tau$ are not captured by the description in terms of the matrices ${\mathbf M}^t$.

The extent to which a given ${\mathbf M}^t$ can be transplanted to a different timepoint $t$ or to describe a system which was initialised with different starting conditions, or with a different history of control parameters, must be judged by the practitioner.  If the series of ${\mathbf M}^t$ is stable for successive timepoints then this indicates that the dynamics, at least with respect to $\lambda$, may have converged to some stationary limit.  Under the assumption of stationarity in the dynamics, diagonalisation of  ${\mathbf M}^t$ to find the infinite-time distribution with respect to $\lambda$  can usefully be carried out, as well as other tactics commonly used to condense a description of the kinetics from a Markov-like matrix of estimated transition probabilities \cite{Wales2009}.

\subsection{S-PRES Algorithm Pseudocode}
\label{sec:pseudocode}

To aid implementation a step-by-step guide to an example program is provided. 

The set of starting configurations can be prepared in any way that is of interest. E.g.~one could prepare an equilibrium set for a given control parameter and then use S-PRES to perform a quench, i.e. change the parameter and observe the system dynamics. Or one could start out from a single configuration and observe how trajectories diverge from this point.   
 If the system is intended to be set up in a stationary state then a conventional rare event sampling method or an initialising round of S-PRES can be run for whatever length of time is needed to prepare a set of configurations with correct associated weights over a good range of $\lambda$.

\begin{enumerate}
\item{Prepare a set of (one or more) configurations of the system at time $t=0$.}
\item{Find $\lambda$ for each configuration, and associate it to the appropriate discrete bin w.r.t. $\lambda$.}
\item{Prepare a vector ${\vec d}^0$ giving the estimated occupation probability of each bin on $\lambda$ at $t=0$.}
\item{Prepare a vector ${\vec n}^0$ giving the number of shots to make from each bin at $t=0$.}
\item{Loop for $t=0$ to $t=\infty$:}
\begin{enumerate}
\item{Set all transition probability estimates $M^t_{i,j} = 0$.}
\item{Loop for all $i$ s.t.~${d}_{i}^t \neq 0$:}
\begin{enumerate}
\item{Repeat $n_{i}^t$ times:}
\begin{enumerate}
\item{Select a configuration in the bin $i$. At $t>0$ use eqn.~\ref{eqn:psel} (with a shift of index) to weight the configurations according to their previous bin. At $t=0$ set all configurations in $i$ as equiprobable.}
\item{Run dynamics of duration $\tau$.}
\item{Calculate $\lambda$ for the evolved configuration and find bin $j$ given $\lambda$.}
\item{Associate the evolved configuration to bin $j$ for the next timestep, also recording its origin as $i$.}
\item{Set $M_{i,j}^t = M_{i,j}^t+{1}/{n_{i}^t}$.}
\end{enumerate}
\item{Set sampling rate ${n_{i}}^{t+1}$ using eqn.~\ref{eqn:sampfreq}.}
\end{enumerate}
\item{Print the matrix ${\mathbf M}^t$ and the vector ${\vec d}^t$.}
\item{Print any further observables derived using eqn.~\ref{eqn:obsweights}.}
\item{Set ${\vec d}^{t+\tau} = {\mathbf M}^t{\vec d}^t$.}
\item{Set $t = t+\tau$.}
\end{enumerate}
\end{enumerate}

\subsection{Boundary Conditions for Flux Calculations }
\label{sec:obs}

S-PRES can be used in two ways, either to calculate the time dependent probability distribution of some static observable or to calculate the time dependent reaction flux $\phi(t)$ between two specifically chosen `source' and `sink' bins on $\lambda$ (which in the following we call $A$ and $B$). The latter quantity is the non-stationary analogue of that which is usually calculated using TPS and FFS methods, and the former of that which is usually calculated via IS techniques. Flux calculations typically require special treatment of boundary conditions, in order to remove the effects of granularity in time and in order to create a system which can remain far from equilibrium indefinitely.

In order to achieve a definition of the forward flux which is strictly independent of $\tau$ it is necessary that runs which enter the `sink' region, $B$, are halted immediately. This may be computationally costly if the coordinate $\lambda$ is costly to calculate, but cannot be avoided if an accurate flux is desired.  If paths were allowed to enter and leave $B$, this would not correspond to the accepted definition of forward reaction flux as the probability per unit time of a first passage from $A$ to $B$.

The region $B$ is treated as absorbing in this way: at the end of each timestep, the probability vector entry corresponding to $B$ is set equal to zero and the entry corresponding to the region $A$ is incremented by the flux which has been deleted. No new configurations are actually transferred to $A$, only some of the `probability mass' tracked by ${\vec d}^t$. This `short circuit' of the matrix is equivalent to a system with an infinite reservoir of states in $A$ and absorbing boundary conditions at $B$; which is the premise normally adopted for FFS.

A second restriction, which should not in general be used but which {\it was} imposed on the FFS-like ``flux" variant of the method for the calculations carried out here, is to instantly terminate any paths which return to $A$; and then re-initialise them with a random configuration from within $A$. This setup describes a slightly unphysical situation, but was required here to achieve exact correspondence of the definition of flux with existing steady-state FFS calculations \cite{Allen2008}, such that only paths from $A$ to $B$ which make the journey in a single pass without any return to $A$ are considered.  

If it is preferred to calculate the time series of the state vector ${\vec d}^t$ and the matrices ${\mathbf M}^t$ (or some other extracted observable) without specific definition of a flux then no special bins $A$ or $B$ are defined.

\subsection{Non-Requirement for Poisson Statistics}

The assumption of Poisson statistics; that rare events occur independently and without correlation; is required by most existing methods \cite{Allen2009}.  This assumption may be an unwelcome limitation and is not required by S-PRES. (Although in the example of (sec.~\ref{sec:shearIsing}) boundary conditions were set up so as to force Poisson behaviour).  The weighted directed acyclic graph (WDAG) of configurations which S-PRES generates can be used to measure the deviation from Poisson statistics.  Define $\overline{P}(X_{t'}|X_t)$ as the estimated conditional probability of event $X$ at time $t'>t$ given that the system also experienced $X$ at time $t$.  This is measured by performing a sum over the weights of configurations at $t'$, counting only that set which are descended directly from configurations which experienced $X$ at time $t$, $\{w_i\}$,  and a second sum over only those which experienced the event at both times, $\{w_i'\}$:

\begin{equation}
\label{eqn:poisson}
\overline{P}(X_{t'}|X_t) = \sum_i \{w_i'\} / \sum_i \{w_i\}
\end{equation}

In this example use of the WDAG, the size of the set of joint events $\{w_i'\}$ may in practice be so small as to cause sampling errors unless the deviation from Poisson statistics is large or a large calculation is carried out.  Storage of the entire WDAG is likely to be cumbersome for many practical applications in which the storage capacity to describe every configuration will quickly mount up.

\section{Application to Rare Events in the Ising Model - Glauber-Ising under Imposed Shear Flow}
\label{sec:shearIsing}

\subsection{Introduction to System}

The example of nucleation in the 2D Ising model on a square lattice under imposed shear flow is 
a case that has been studied (although only for constant shear rate) using FFS \cite{Allen2006a,Allen2008};
allowing for direct comparison of our results with those from an established method.  This simple model exhibits rich non-equilibrium phase behaviour \cite{Cirillo2005}, however in this instance it is employed only to demonstrate the use of the sampling algorithm.

The system was set up as follows (duplicating the FFS studies): Glauber dynamics were used to evolve the spins at each lattice site, meaning that at each `sweep' $L \times L$ sites were chosen randomly to have their spins reassigned according to a Boltzmann-weighted probability. The system was prepared with all ($65 \times 65$) spins down. A weak upward external field was applied, rendering the prepared state metastable relative to the stable state in which all or most spins are up.  Shear flow can either accelerate or retard the formation of a nucleus of up-spins and the transition to the stable state, depending on the shear flow rate \cite{Allen2006a}.  

Shear flow was applied $L$ times at each sweep by randomly selecting one of the $L$ horizontal lines between rows of lattice sites and applying a move with probability $\dot{\gamma}$ such that all sites above it are translated one space to the left, with periodic wrapping such that the spin at $i=1$ becomes the spin at $i=L$; this is a simple model of infinite two-dimensional laminar Couette flow.  

A subtlety enters in the treatment of the vertical periodic boundaries (the interaction between sites with $j=1$ and $j=L$): in order to avoid shearing along this line unless it has been explicitly selected, an offset pointer is maintained so that even after the spins in the row with $j=1$ have been moved (after a shear at say, the boundary between $j=5$ and $j=6$) they remain in contact via periodic imaging with the same spins in the row $j=N$ as they were before.    

A long discussion of the detailed implementation of this model system is available \cite{Allen2008}.  The notation $\dot{\gamma}$ to indicate the rate of the imposed shear flow is used here for consistency with this earlier work and has no relation to the $\gamma$ of eqn.~\ref{eqn:sampfreq}, which is used to indicate the `damping' constant applied to stabilise sampling rates.  

In order to generalise the steady-state model having constant shear flow to a simple and directly comparable non-steady-state case we subjected the system to a time-series of three different shear flow rates within the low-shear (nucleation-enhancing) regime, allowing the nucleation rate to relax to the steady-state value after each change of shear flow rate. In order to produce directly comparable data to the FFS studies, the time dependent forward flux for the system was calculated using the same system parameters described for the steady state calculations in refs.~\cite{Allen2006a,Allen2008}; which is to say size $L=65$, coupling constant $J=0.65k_BT$ and external field strength $h=0.05k_BT$.  At $\dot{\gamma}=0$  the system can be considered to be in a state of quasi-equilibrium where a meta-stable basin and stable basin are separated by a large free energy barrier. Classical Nucleation Theory (CNT) gives the size of the barrier to nucleation as $\approx 22k_BT$ for this regime \cite{Sear2006},  signifying that a rare-event technique is strongly recommended to extract meaningful statistics by simulation.

\subsection{Definition of Coordinate Bins for Shear-Ising Calculation}

The coordinate over configurations was defined simply as the total number of up spins present, $\lambda=N_{up}$. The source bin, $A$, was defined as $\lambda < 25$ and the sink bin $B$ was defined as  $\lambda > 2005$ (as in the FFS calculation \cite{Allen2008}). The intervening space on $\lambda$ was divided into $990$ equal increments.  It might have been possible to make a more sympathetic definition of the intervening bins, such as by spacing them more closely together for smaller values of $\lambda$ where the dynamics on $\lambda$ is expected to be slow, however this crude binning was found to be effective.  

The sampling parameters $\gamma=0.5$, $\tau=10$ and $N=100$ were used. Occasional `dead-end' configurations manifested, where $\lambda$ was large due to multiple isolated clusters of spins rather than due to a single nucleus: a maximum $n_i$ threshold of $2000$ was therefore set, in order to prevent eqn.~\ref{eqn:sampfreq} from diverging due to these instances.

\subsection{Results for Shear-Ising Calculation}

\begin{figure}
\includegraphics[width=0.5\textwidth]{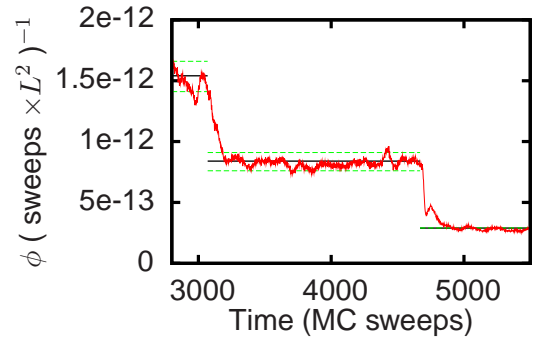}
\caption{\label{fig:exampleUses_shi} {\bf Example use of S-PRES:  Nucleation in the 2D Ising model under shear.} Solid horizontal lines: reference steady state nucleation rates for each value of imposed shear \cite{Allen2006a} (dashed lines show the reported errorbars). Fluctuating trace (red online): S-PRES time-series as the shear is changed. }
\end{figure}

It was necessary to run the calculation for $2500$ MC sweeps at the initial shear rate $\dot{\gamma}=0.04$ before all bins of the coordinate $\lambda$ were populated, allowing meaningful statistics to be collected.  Fig.~\ref{fig:exampleUses_shi} shows the flux against time as the shear rates were changed ($\dot{\gamma}=0.04$, $0.02$, and $0.0$).  Horizontal lines (black) indicate steady state FFS data from a separate research group \cite{Allen2006a}; the trace (red online) is the S-PRES results. After each change of shear rate the time-dependent flux relaxes to the known steady state value (actually the quasi-equilibrium value in the case $\dot{\gamma}=0.0$), validating the method. The trace is an average over $100$ independent runs.

\section{Application to Rare Events in the Ising Model - Kawasaki-Ising after a Quench}

\subsection{Introduction to System}

Phase separation in the 2D Ising model after a quench into the temperature region between the nucleation-dominated and spinodal decomposition-dominated regimes is a quintessential problem in non-equilbirium dynamics.  Under Kawasaki dynamics (sometimes called a `lattice gas') the total magnetisation is conserved and time-evolution is controlled by diffusion of spins. The base timescale of the system is set by one MC sweep, equal to $N_{up}$ attempts to move a random up-spin.  Because the diffusive and evaporative behaviour of spin clusters is determined by both their size and shape it is difficult to predict their rates of collision and growth or shrinkage and the evolution of the size distribution of clusters over time.  Despite these difficulties a theory based on the iterative evolution of a population vector of clusters of different sizes, $p_n(t)$ is available from the literature \cite{Mirold1977}, which has not until now received direct validation from simulation studies (although a closely related approach has had the benefit of such scrutiny \cite{Penrose1983}).  The existence of an untested theory for such a simple but important model system is an ideal opportunity to further demonstrate the S-PRES method while at the same time making a small contribution to the basic study of phase-change dynamics.

 The equilibrium thermodynamics of this model are well understood, as are the phase-change dynamics in both the nucleation-dominated (surface-energy limited) and spinodal decomposition-dominated (diffusion limited) regimes \cite{Puri2009}.  We carried out an instantaneous quench from $T=\infty$ to $T=0.6T_c$, which lies between these two regimes,  for a system of $100\times100$ spins with a $0.1$ concentration of up-spins.  The coupling constant $J$ was set to $1k_BT$.  $T=0.6T_c$ was set to $1.36151$, using the Onsager result of $T_c=2/ln(1+\sqrt{2})$ \cite{Onsager1944}.

\subsection{Definition of Coordinate Bins for Kawasaki-Ising after a Quench}

As the coordinate we chose $\lambda=\sum_{c}(n_c - 1)$, where $n_c$ is the number of spins in cluster $c$ and a cluster is defined as a connected group of up spins. This coordinate was chosen because it is simple to calculate and has a value of zero when all up spins are isolated, increasing after any collision between spins or clusters.   The lowest bin was defined as $\lambda \le 9$ and the highest bin was defined as $\lambda \ge 91$, with the intervening integer values assigned to a single bin each.  

The sampling parameters $\gamma=0.5$, $\tau=10$ and $N=100$ were used. 

\subsection{Results for Kawasaki-Ising Calculation}

In order to prepare initial states including unusually large clusters the system was prepared in the $T=\infty$ regime using some few hundreds of iterations of S-PRES in order to achieve statistics down into the range $p_n=10^{-12}$ before applying the quench.  

\begin{figure}
\includegraphics[width=0.45\textwidth]{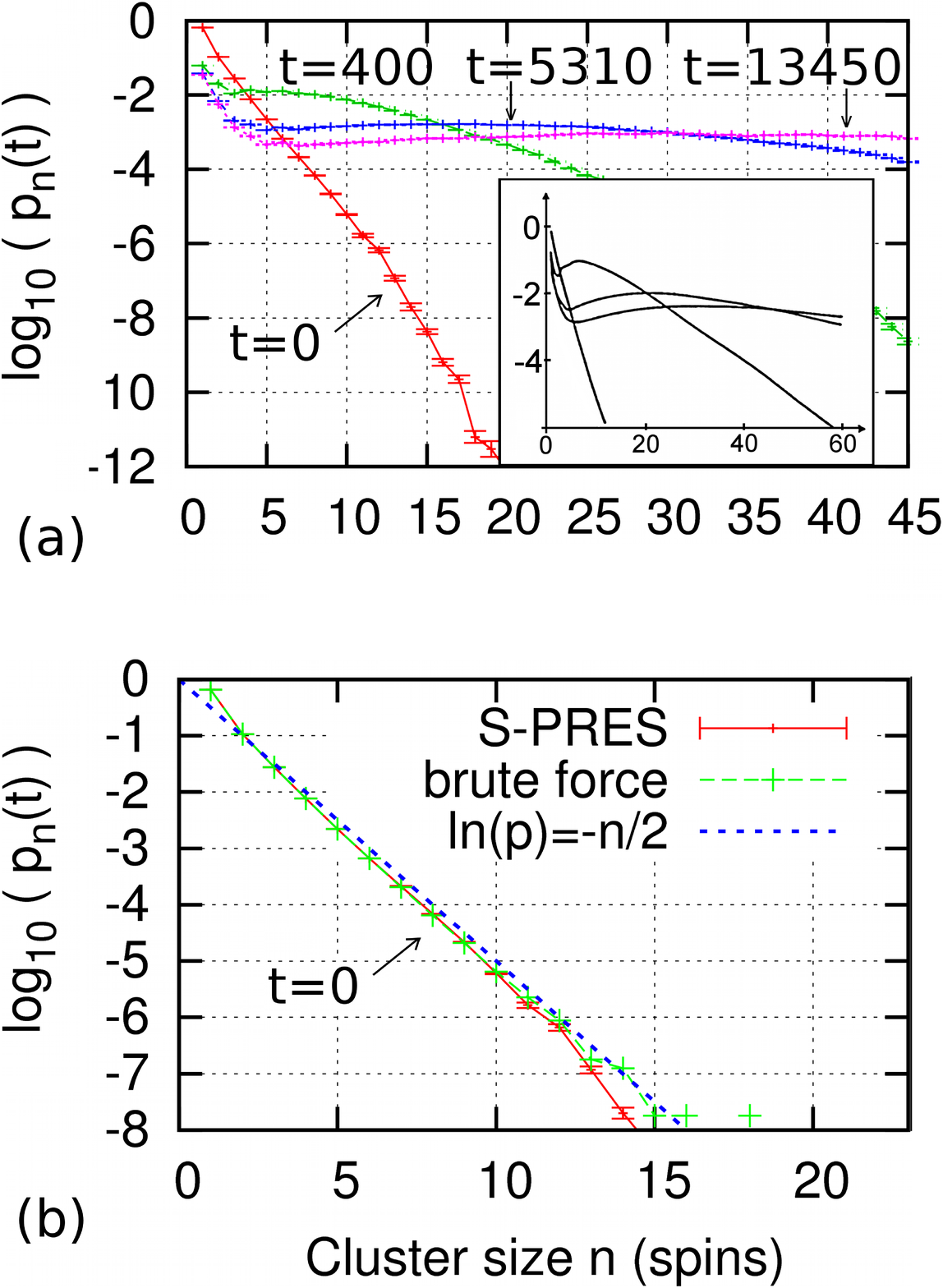}
\caption{\label{fig:exampleUses_kaw} {\bf (a) Example use of S-PRES: Temperature quench of the 2D Ising model with conserved order parameter.} Calculated time evolution of the domain-size distribution is compared with theory \cite{Mirold1977} (inset).  The theory is well outside the error bars (which are invisible except for very small $p_n$), but does provides qualitative agreement in so far as reproducing the shapes of the four curves.  {\bf (b) Special case $\mathbf{t=0,T=\infty}$.} At $T=\infty$ a brute-force calculation is easy, so is superimposed on the S-PRES data down to $p_n\approx 10^{-8}$. The standard result $p(n)=e^{-n/2}$ is also shown. {\it Colour online.}}
\end{figure}

The probability distribution of cluster sizes $\vec{p_n}$ is not directly available from the probability distribution of reaction coordinate bins $\vec{d}$; instead it was required to calculate it as an average over all configurations generated at each timestep, weighted according to eqn.~\ref{eqn:obsweights}.  In fig.~\ref{fig:exampleUses_kaw} we show S-PRES results for $\vec{p_n}$.  There is good qualitative agreement with the theory, which is shown in fig.~\ref{fig:exampleUses_kaw}-inset.  The results shown are averaged over $10$ independent calculations; error bars are the estimated standard errors over the $10$ values. Brute force calculation in the $T=\infty$ regime is very cheap due to the lack of interactions between spins at this temperature, therefore data at $t=0,T=\infty$ from a brute force calculation separate to the S-PRES calculation is also shown in fig.~\ref{fig:exampleUses_kaw} (and also the standard result $p_n=e^{-n/2}$ \cite{Stauffer1979}).  The existing quantitative results for the infinite temperature case highlight a potential source of problems caused by the error behaviour of S-PRES - until convergence is achieved, probabilities of rare states are reported as zero; meaning that S-PRES will converge on the correct values from below.  

To comprehensively explore the applicability of the Mirold-Binder theory is not the aim of this work, and would require further data over a wide range of temperatures and concentrations, however to provide numerical results for systems previously accessible only to theory is an example of the type of research into phase change dynamics which can be carried out using S-PRES. 

\section{Application to Rare Events in a Time-Dependent Asymmetric Exclusion Process}
\subsection{Introduction to System}

An Asymmetric Exclusion Process (ASEP) is a simple model for driven stochastic transport.  Here we discuss the ``parallel-open'' (p-o) ASEP, as has been characterised by Sch\"utz \cite{Schutz1993}.  In this model, particles are introduced at the origin with a probability $\alpha$ at each sweep; and removed from the right boundary with probability $\beta$.  Between the two boundaries, a deterministic update rule allows particles to move from left to right providing that a vacancy exists.  When $\alpha = \beta$, the system becomes critical, with a divergent correlation length. Particles queue up at the right boundary of the system, forming a block with density $(1-\beta)$; and the remainder of the system has fast-moving traffic with density $\alpha$.  The phase boundary moves stochastically in the critical state according to a random walk.

\subsection{Simulation Setup}

In order to validate S-PRES against the quite tractable time-dependent properties of the ASEP, the system was initialised without any particles; and allowed to gradually approach the steady state, in analogy to the morning traffic along a busy road.  The length $L$ was set as $500$ sites and the parameters $\alpha$ and $\beta$ were both set as $0.01$.  12,000 brute force calculations were run, each of duration $10^7$ sweeps.  A single S-PRES calculation was also set up, with $\lambda$ defined as the number of particles, divided over 100 equal-sized bins. The S-PRES parameters $\tau=500$ and $\gamma=0.5$ were used.

\subsection{Results for ASEP Calculation}

The rare event in this case is the full occupation of the system, i.e.~the 
particle number equalling the number of sites.
Statistics were collected from the S-PRES and brute force runs on the probability of the rare event $P(full)$.  This is available in the steady-state limit from \cite{Schutz1993} as $P(full)_{t\rightarrow\infty}=2(1-\alpha)^{L/2}/L$.  Assuming that the phase boundary moves as a random walk starting from the origin at $t=0$, and that the density of the dense phase is constant; then the time-dependent value of $P(full)$ is readily available by numerically iterating Fick's first law, beginning with the probability density defined as $1.0$ at the origin and $0$ elsewhere. 

\begin{figure}
\includegraphics[width=0.45\textwidth]{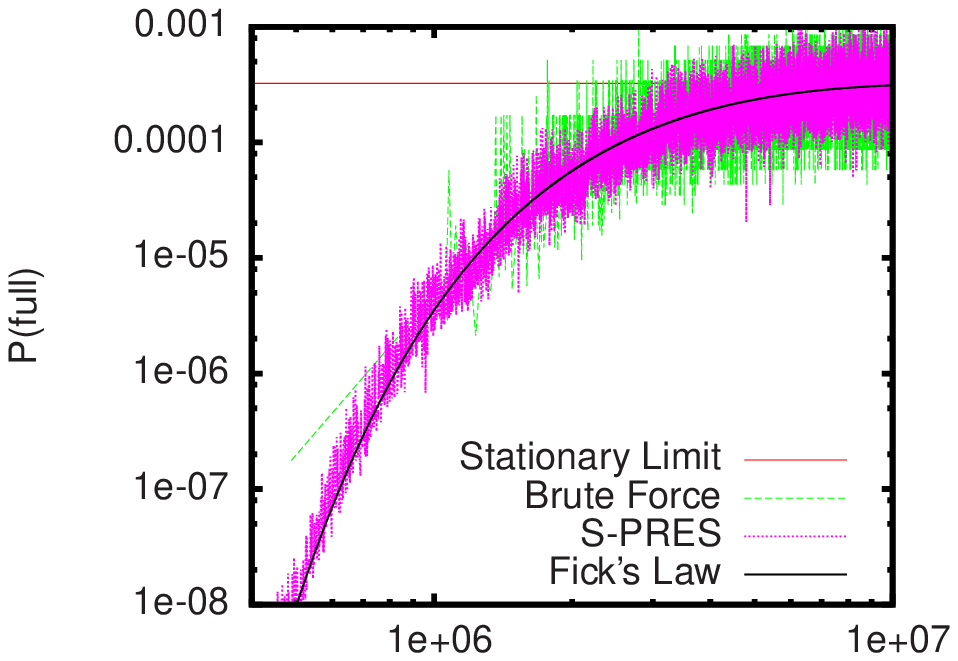}
\includegraphics[width=0.45\textwidth]{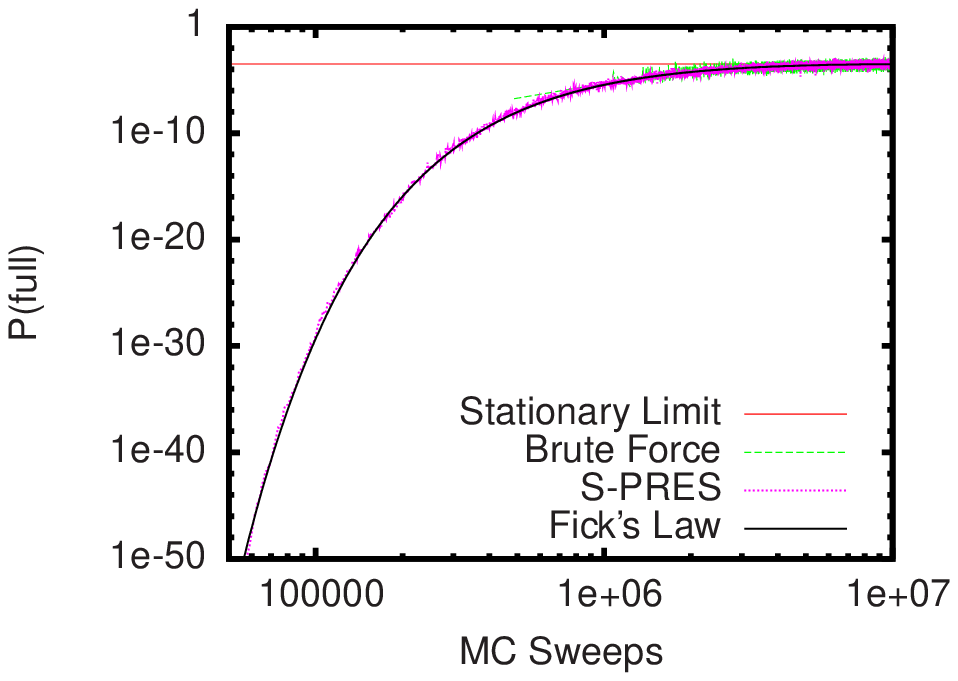}
\caption{\label{fig:asep} {\bf Probability that the ASEP is completely full}, given that it is empty at $t=0$. The S-PRES calculation agrees with theory over 46 decades. {\it Colour online.}}
\end{figure}

The S-PRES and brute force calculations converged to the steady-state limit and the S-PRES calculation was also able to confirm that kinetic properties were very accurately predicted by the assumption of Fickian diffusion of the phase boundary.  This is shown in fig.~\ref{fig:asep}.

\section{Example Use of the Matrices $\mathbf{M^t}$}
\label{sec:committor}

The structure of $\mathbf{M}$ with respect to time can be analysed in order to estimate the usefulness of the coordinate projection which has been employed and to pursue insight into the mechanics of the system under consideration.  An example is the extraction of committor probabilities.

The `committor probability' $p_B(s)$ or the ultimate probability that a given configuration $s$ will complete the reaction before returning to some initial state is a quantity generally of interest in the analysis of rare events.  In a non-stationary system this value can change with respect to time.  An easy estimate of committor probabilities at a given time $t_0$ can be achieved by using the time series of $\mathbf{M}$ in the following procedure:

\begin{enumerate}
\item{Create two `sink' bins  $A$ and $B$ s.t.~$\forall t > t_0$: $M^t_{A,A}=1$, $M^t_{B,B}=1$, $M^t_{A,j\neq A}=0$, $M^t_{B,j\neq B}=0$.}
\item{Repeat for each bin $b \notin \{A,B\}$}
\begin{enumerate}
\item{Initialise a vector $\vec{d}$ s.t.~$d_b=1$ and $d_{i}=0$ $\forall i \ne b$.}
\item{Apply the time series of modified $\mathbf{M^{t}}$ to each $\vec{d}$, beginning at $t=t_0$ until $d_i\approx 0$ $\forall i \notin \{A,B\}$.}
\item{$d_B$ now holds the expected value of $p_B(s)$ over the bin $b$.}
\end{enumerate}
\end{enumerate}

The procedure above gives the committor only with respect to the bins on the projected coordinate $\lambda$; the main purpose of such an analysis is to evaluate the usefulness of the particular definition of $\lambda$.  It is considered that the closest possible identity between $\lambda$ and the committor probability gives the most efficiently enhanced sampling for rare event methods \cite{Ma2005}. If $\lambda$ does not determine $p_B$ or if the granularity of the binning is large near to sharp changes in $p_B$, then S-PRES becomes less useful and alternative strategies might be required. If an initial rough calculation can be made to work, then it is possible to record online (without the assumption of mixing) the mapping between $\lambda$ and some different observable using eqn.~\ref{eqn:obsweights}.  In the case that large computer memory is available then the entire WDAG of configurations connected by path segments can be stored; allowing formal methods of projection onto subspaces of manageable dimensionality \cite{Amadei1993,Yan2005a} to be experimented with offline.  

\begin{figure}
\includegraphics[width=0.5\textwidth]{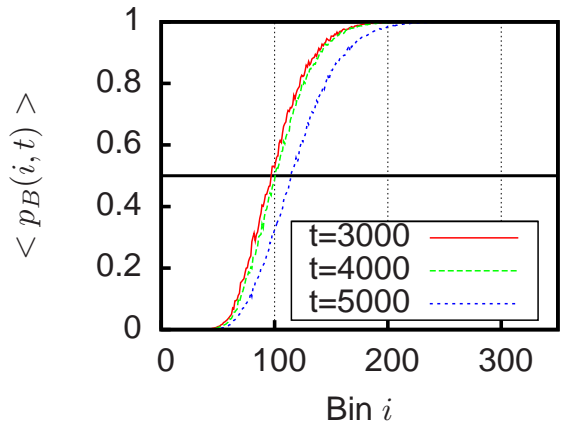}
\caption{\label{fig:committors} {\bf Committor probabilities for 2D-Ising under shear}, assuming that the system is initialised in bin $i$ at each time $t=3000;4000;5000$ (corresponding to $\dot{\gamma}=0.04$, $0.02$, and $0.0$). The committor $p_B(i)$ moves to higher bins for later timesteps (lower shears), which is consistent with the lower nucleation rates observed. {\it Colour online.}}
\end{figure}

Fig.~\ref{fig:committors} shows the committor probability distributions for three timepoints in the evolution of the variable-shear system of (sec.~\ref{sec:shearIsing}) corresponding to three different shear-rates.  The shifting of the distribution to the right for lower shears is consistent with the smaller reaction flux observed. This effect is due to the fact that the frequency of cluster collisions decreases with decreasing $\dot{\gamma}$ more quickly than does the ``cluster evaporation'' rate.

Further statistics of interest for a typical system might include the transmission coefficient $\kappa$ or the width of the committor distribution, as discussed for the equilibrium 3D Ising model in ref.~\cite{Pan2004}.  These statistics can be calculated with respect to $\lambda$ as above; or with respect to an arbitrary variable by using the WDAG.

%\subsection{Relation to the Master Equation}

%A Chemical Master Equation (CME) can be defined as the square matrix of forward and backward reaction rates between distinct chemical species in a well-mixed solution.  Diagonalisation of this matrix gives the possible stable and metastable concentration profiles over the various species; as well as quick access to descriptions of the kinetics under the (probably false) assumption of a Markov process \cite{Wales2009}; making it a very useful representation. 

%The important difference between the matrices $\mathbf{M^t}$ discussed in this work and CME matrices is that they describe transitions between arbitrary subsets of the configurations of any system rather than strictly referring to chemical concentrations.  They are time-dependent; and do not in general have the Markov property, meaning that arbitrary initialisations such as that used for the committor estimation of (sec.~\ref{sec:committor}) are not guaranteed to accurately predict the dynamics for specific configurations.  A valuable discussion of the relationship between stochastic (microscopic) simulation and continuous and deterministic (macroscopic) calculations using the CME to evolve a vector of concentrations is laid out in a relatively recent review article \cite{Gillespie2007}.  

\section{Scaling and Choice of Binning}

\subsection{Scaling with Relation to Fineness of Binning}

If the S-PRES calculation is set up in order to find rare states, then it 
has two phases.  In the first phase (`population') the goal is to achieve a state whereby one or more configurations are associated with each bin, allowing rare events to be observed.  In the second phase (`observation') the goal is to continue the dynamics and observe the time-evolving behaviour.  To make a loose scaling argument from equilibrium statistical mechanics, if we assume that the number of bins  $N_B$ is large enough that no significant free energy barriers exist between adjacent bins, but that only one new bin is populated at each iteration of the algorithm, then the total time required for the population phase should be roughly proportional to $N_B^2/2$.  During the observation phase, the time required per iteration should continue to be linear with the number of bins.  Therefore the population phase is considered as the performance bottleneck of the method and  `speedup' is defined as the expected number of MC sweeps needed to observe the first rare event using brute force (equal to $L^2/\phi$ for the shear-Ising example system) divided by the expected number of MC sweeps needed to observe the first rare event using S-PRES, which marks the completion of the population phase.

\begin{figure}
\includegraphics[width=0.5\textwidth]{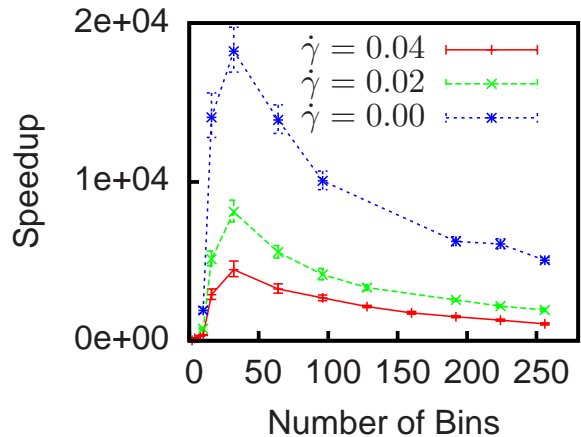}
\caption{\label{fig:benchMark} {\bf Scaling of algorithm efficiency:} performance is more robust to excessive numbers of phase space bins than it is to insufficient numbers of bins.  As the probability of the rare event in consideration decreases (smaller $\phi$) the speedup becomes proportionally more. Each trace is an average over 10 independent calculations. {\it Colour online.}}
\end{figure}

We repeated the shear calculations of (sec.~\ref{sec:shearIsing}) at constant $\dot{\gamma} = 0.04,0.02$ and $0.0$ and with $L=50$; for various numbers of equally spaced bins, even for small numbers of bins where the assumption of closely spaced bins no longer holds. Fig.~\ref{fig:benchMark} shows speedup versus the number of bins for calculations at the three different shear rates.  At the shear rates for which nucleation is more rare, the speedup is proportionally greater. The speedup was robust to the use of excessive numbers of bins.  The smaller system of $L=50$ was chosen for the benchmarking because it is computationally less expensive; reaction fluxes were roughly equal to those observed for the larger system, with $\phi=1.5\times10^{-12},0.8\times10^{-12}$ and $0.3\times10^{-12}$ for the three shear rates.

\subsection{Robustness to Non-Monotonic Binning}

A common thought experiment used to test coordinate-based rare event methods such as FFS, umbrella sampling or S-PRES is to imagine a system for which the the projected coordinate $\lambda$ is non-monotonic or sometimes orthogonal with respect to the true reaction coordinate \cite{Allen2009,Dickson2009}.  A practical example of this is protein folding, where even simple proteins and peptides can move through sequences of transition states which are dissimilar to each other and to both the unfolded and folded conformations \cite{Bolhuis2000,Bartlett2010}, making it difficult to define a useful projection of the progress of the reaction without detailed prior knowledge of the the folding mechanism.  

S-PRES is robust to this situation in the sense that S-shaped trajectories can be developed by the algorithm because paths which move backward as well as forward in $\lambda$ are generated and stored; however it is still better to choose coordinate projections which are near-monotonic with respect to the real progress of the reaction because any bins which represent multiple stages of the `true' reaction coordinate will require larger populations in order to give stable sampling; which will need to be crudely dealt with by setting a small $\gamma$ and large $N$ in eqn.~\ref{eqn:sampfreq}.

\section{Concluding Remarks}

This paper presents a method, S-PRES, to investigate rare events in non-equilbrium 
and non-steady-state dynamics. S-PRES can compute the evolution of the probabilities of 
rare events or rare states in any stochastic system with respect to time, 
providing that a suitable binning on the phase space can be defined. 
The method is based on Forward Flux Sampling with modifications to permit tracking of the ages of the configurations sampled. A version of the pruned-enriched Rosenbluth method is applied 
to the generation of path segments in order to achieve efficient sampling. 

To demonstrate the method we calculated phase change kinetics in the Ising model both under shear and after a temperature quench; confirming existing results from theory and simulation. We also confirmed theoretical results for the time-dependent critical behaviour of a model of driven diffusive transport. We anticipate that the method is useful for a very wide range of time-evolving processes in nature. Possibilities include the probabilities of abnormal cell differentiation during embryogenesis, fracture nucleation in materials under impact or time-varying load, and nucleation in glassy materials approaching dynamic arrest.

\begin{acknowledgments}
We thank Kurt Binder, Martin Weigel, Colin Fox and Giovanni Peccati for discussions.
This work was supported by the DFG (Emmy Noether Programme and SPP1296) and with CPU time by NIC J\"ulich and by the UK National Grid Service.
\end{acknowledgments}

%\bibliography{ssmmBiblio} 
%\end{document}

%Merlin.mbs v4.21 2009-07-09.
%
\end{document}